\documentclass[letterpaper, 10 pt, conference]{ieeeconf}  

\IEEEoverridecommandlockouts                              

\usepackage{graphicx} 
\usepackage{amsmath}
\usepackage{amsfonts}
\usepackage{amssymb}
\usepackage{mathtools}
\usepackage[dvipsnames]{xcolor}
\usepackage{bm}
\usepackage{cite}
\usepackage{enumerate}

\newcommand{\KP}{\ensuremath{(K,P)}}
\newcommand{\KPk}{\ensuremath{(K_k,P_k)}}
\newcommand{\setKD}{\ensuremath{\mathcal{K}_{\mathcal{D}}}}
\newcommand{\setD}{\ensuremath{\mathcal{D}}}

\newcommand{\probLMI}[2]{\mathbf{P}(x_k,#1,#2)}
\newcommand{\SigmaV}{\Sigma^{\mathcal{V}}}

\newcommand{\argmin}{\arg\min}

\newtheorem{definition}{Definition}
\newtheorem{remark}{Remark}

\newtheorem{lemma}{Lemma}
\newtheorem{assumption}{Assumption}

\newtheorem{thm}{Theorem}
\usepackage{standalone}
\usepackage{tikz}
\usepackage{pgfplots}
\pgfplotsset{compat=newest}
\pgfplotsset{plot coordinates/math parser=false}
\usepgfplotslibrary{statistics}

\title{\LARGE \bf An adaptive extension to robust data-driven predictive control\\under parametric uncertainty}
\author{Ignacio Sanchez, Filiberto Fele and Daniel Limon
\thanks{Support from grants PID2022-142946NA-I00 and PID2022-141159OB-I00 funded by MICIU/AEI/ 10.13039/501100011033 and by ERDF/EU is gratefully acknowledged. F.~Fele also acknowledges support from
grant RYC2021-033960-I funded by MICIU/AEI/ 10.13039/501100011033 and European Union NextGenerationEU/PRTR.}
\thanks{The authors are with the Dept.~of Systems Engineering and Automation, 
          University of Seville, Av.~de los Descubrimientos s/n, 41092 Seville, Spain.
         {\tt\small \{isanchez10, ffele, dlm\}@us.es}}}%

\begin{document}
\maketitle
\thispagestyle{empty}
\pagestyle{empty}

\renewcommand{\arraystretch}{0.9}
\setlength{\arraycolsep}{3pt}  

\begin{abstract}
Robust data-driven controllers typically rely on datasets from previous experiments, which embed information on the variability of the system parameters across past operational conditions.
Complementarily, data collected online can contribute to improving the feedback performance relative to the current system's conditions, but are unable to account for the overall---possibly time-varying---system operation.

With this in mind, we consider the problem of stabilizing a time-varying linear system, whose parameters are only known to lie within a bounded polytopic set. Taking a robust data-driven approach, we synthesize the control law by simultaneously leveraging two sets of historical state and input measures: an \emph{offline} dataset---which covers the extreme variations of the system parameters---and an \emph{online} dataset consisting of a rolling window of the latest state and input samples.

Our approach relies on the \emph{data informativity} framework, allowing a direct data-to-feedback design based on standard Lyapunov arguments. The procedure is implemented via semi-definite optimization: this also yields an upper bound on the cost-to-go for the class of systems that are consistent with the online data, while guaranteeing a decreasing cost for all systems compatible with the offline data.
Numerical experiments are presented to illustrate the effectiveness of the proposed controller.
\end{abstract}

\section{Introduction}
The data-driven control paradigm relies on measured system trajectories for the synthesis of a feedback control law, either through an \textit{indirect} approach---involving model identification---or by extracting the control law  straight from the available data, in what is referred to as \textit{direct} approach~\cite{markovsky2021behavioral}.
Lately, large research efforts have been devoted to translate classic modelling approaches---typically based on some combination of model identification or derivation from first principles---into direct data-driven designs which bypass the modelling and identification processes; fundamental formulas and related results were presented in the context of optimal and robust control in~\cite{de2019formulas}.

This paradigm was shown to fit particularly well the model predictive control (MPC) framework~\cite{CamachoBordonsBOOK,rawlings2020model}, giving rise to \emph{data-driven MPC} formulations~\cite{Berberich2021Data, coulson2019data}.
Many such formulations build upon the Willem's Fundamental Lemma and, as such, are sensibly dependent on trajectory datasets satisfying conditions of \emph{persistent excitation}~\cite{willems2005note}.

More recently, the design of controllers under less restrictive conditions was made possible by exploiting the \emph{data informativity} notion~\cite{van2020data,van2021matrix}.
This enables the design of controllers  when the available data is not descriptive enough to uniquely identify some dynamics.
In particular, the data is said to be informative for some system property (e.g., identification, stabilization) if such property can be derived from the available data. It was shown in~\cite{van2020willems} that data informativity requirements for identification are stricter than those necessary for stabilization.
Therefore, one could determine the existence of a stabilizing controller for an entire class of systems that are consistent with the data, while being unable to discern which specific system in that class has effectively generated the data (i.e., the data may be informative for stabilization but not for identification).

This framework has been successfully used in the design of robust data-driven MPC formulations: 
building upon the classic formulation of min-max 
MPC for systems described by Linear Differential Inclusions (LDI) with polytopically bounded parametric uncertainty introduced in~\cite{kothare1996robust}, the authors of~\cite{nguyen2023lmi,HuLiu_L-CSS24} proposed data-driven formulations which make use of input-state measures collected offline. 
The authors of~\cite{xie2026data} studied the stabilization problem for linear time-invariant systems under additive noise, while the uncertainty about the system is reduced during operation by processing an increasingly larger dataset at each sampling time.
A robust data-driven controller formulation is presented in~\cite{seuret2024robust} for the stabilization of an input-saturated system subject to additive disturbances.
In~\cite{PORTILLA2025231}, unknown input-output delays are considered in the  design of a data-driven stabilizing controller for an unknown linear system. 
It is important to note that all these controllers are designed in terms of the data-informativity framework and are implemented using Linear Matrix Inequalities (LMI), which result in tractable optimization problems and for which efficient solvers are available.
Also, in these works the informativity for stabilization is assessed in terms of the Finsler's lemma~\cite{van2021matrix} or the matrix S-lemma~\cite{vanWaarde2022nonconservative}, that allow certifying positive-definiteness conditions on matrices involving the candidate generator models for the data.

Here, we seek an improvement of the closed-loop performance by enriching the offline dataset---containing a robust description of the LDI system dynamics (namely, vertex realizations of some given polytopic parametric uncertainty)---with new measurements collected online which yield a more precise description of the system around its current operating region.
We derive an adaptive data-driven predictive controller that can stabilize systems with uncertain and possibly time-varying dynamics. When the parametric variation is piecewise-constant (with sufficient dwell time), the produced control law allows to upper bound the cost-to-go for the class of systems consistent with the online data, while it guarantees a decreasing cost for all systems compatible with the offline data.
By incorporating online data through a rolling window of fixed length, our approach enjoys a fixed numerical complexity (as opposed to related proposals in the  literature).
We show across a set of numerical simulations how the proposed controller can achieve better performance compared to existing robust data-driven approaches.

It is worth emphasizing that, in contrast to~\cite{xie2026data}---concerned with the stabilization of an unknown linear system subject to bounded additive noise---we contemplate here parametric uncertainties through an LDI formulation. 
Also, \cite{nguyen2023lmi, HuLiu_L-CSS24} seek to minimize the worst-case control cost against any possible (time-varying) parametric realization within the polytopic bounds obtained from data. Differently from these works, we use a dual prediction scheme, by which we adaptively optimize the performance only for the system that is consistent with the latest data, while still guaranteeing stability for the entire LDI.

\textit{Paper organization:} Section~\ref{sec:problem} sets the context for the data-driven problem formulation.
Sections~\ref{sec:data-informativity} and~\ref{sec:data-informativity-lqr} introduce technical results based on the notion of data informativity for robust stabilization and linear quadratic regulation, which we leverage for our main result, presented in Section~\ref{sec:robust}.
In Section~\ref{sec:numerical}, we examine the performance of the proposed adaptive formulation through simulations.
Section~\ref{sec:discussion} concludes the manuscript.

\section{Preliminaries}\label{sec:problem}
We consider the problem of stabilizing a dynamical system described by the following LDI
\begin{equation}\label{eq:sys_lpv}
\begin{split}
    x_{k+1} &\in \mathbb{S}(x_k,u_k,\mathcal{V}), \\
    \mathbb{S}(x,u,\mathcal{V}) &\coloneqq \{A x + B u \colon (A,B) \in \mathit{Co}(\mathcal{V})\},
\end{split}
\end{equation}
where $x_k \in \mathbb{R}^n$ and $u_k \in \mathbb{R}^m$ denote the state and the input at time $k$, respectively.
The matrices $(A, B)$ are allowed to be time-varying, and are only known to belong to the polytopic set defined by the vertices $\mathcal{V} \coloneqq \{(A^v,B^v) \colon v = 1,\dots,n_v\}$. 

We consider polytopic state and input constraints described as follows:%
\begin{subequations}\label{eq:constraints}
\begin{align}
    \mathcal{X} &\coloneqq \{x \in \mathbb{R}^n : W_x x \leq {1}_r\},\label{eq:constrX} \\
    \mathcal{U} &\coloneqq \{u \in \mathbb{R}^m : W_u u \leq {1}_\ell\}.\label{eq:constrU}
\end{align}
\end{subequations}

\subsection{Data-driven approach}
Throughout this work, $(A^v,B^v)$, $v=1,\ldots,n_v$, are assumed to be unknown. Instead, we take a data-driven approach and rely on the availability of records of system trajectories about each of the vertices in $\mathcal{V}$ (corresponding to, e.g., prescribed limit conditions for operation).
In other words, we assume each vertex in $\mathcal{V}$ is associated with a dataset $\mathcal{D}^v$ collecting $T^v$ triplets of the form $(x_{k+1},x_k,u_k)$. These can be arranged in matrix form as
\begin{align*}
{X}^{v}_{+} &= [{x}^{v}_1, {x}^{v}_2,\dots,  {x}^{v}_{T^{v}}]\in\mathbb{R}^{n \times T^v},\\
    {X}^{v}_{-} &= [{x}^{v}_0, {x}^{v}_1,\dots,  {x}^{v}_{T^{v}-1}] \in \mathbb{R}^{n \times T^v},\\
    {U}^{v}_{-} &= [{u}^{v}_0, {u}^{v}_1,\dots,  {u}^{v}_{T^{v}-1}] \in \mathbb{R}^{m \times T^v},    
\end{align*}
and we define $\mathcal{D}^v \coloneqq ({X}^{v}_{+},{X}^{v}_{-},{U}^{v}_{-})$ for $v = 1, \dots, {n_v}$.

Moreover, we leverage an online dataset $\mathcal{D}^o$  updated at each time $k$ to hold the sequence of the latest $T^o$ triplets. A similar definition follows:
\begin{align*}
    {X}^{o}_{+} &= [{x}^{o}_{k-T^o+1}, {x}^{o}_{k-T^o+2},\dots,  {x}^{o}_k],\\
    {X}^{o}_{-} &= [{x}^{o}_{k-T^o}, {x}^{o}_{k-T^o+1},\dots,  {x}^{o}_{k-1}],\\
    {U}^{o}_{-} &= [{u}^{o}_{k-T^o}, {u}^{o}_{k-T^o+1},\dots,  {u}^{o}_{k-1}],
\end{align*}
and $\mathcal{D}^o \coloneqq (X^o_+,{X}^{o}_{-},{U}^{o}_{-})$.

We note that for any given $k$, the triplet $(x_{k+1},x_k,u_k)$ encodes information in the form of an affine space containing all $(A,B)$ satisfying the relation ${x}_{k+1} = A {x}_k + B {u}_k$. Intersecting all such spaces for all triplets in a dataset yields the following definition.
\begin{definition} \label{defn:consistency}
The class of systems \emph{consistent} with the dataset $\mathcal{D}^v = (X_+^v,X_-^v,U_-^v)$ is
\begin{equation}\label{eq:consistency-characterization}
\Sigma^v \coloneqq \left\{(A,B)\colon
\begin{bmatrix}
    I & A & B
\end{bmatrix} \begin{bmatrix}
    {X}^{v}_+ \\-{X}^{v}_-\\-{U}^{v}_-
\end{bmatrix} = {0}\right\}.
\end{equation}
\end{definition}
\vspace{6pt}

In the following, we will denote by $\mathcal{D}= \{\mathcal{D}^v\}_{v=1}^{n_v}$ the dataset including all (offline) vertex experiments.
Also, $\SigmaV \coloneqq \bigcup_{v=1}^{n_v} \Sigma^v$ is the set of all systems consistent with the offline data~$\mathcal{D}$. 
We assume $\Sigma^v$ is nonempty and $(A^v,B^v)\in\Sigma^v$, for each $v$.

The systems compatible with $\mathcal{D}^o$ constitute the class $\Sigma^o$, according to Definition~\ref{defn:consistency}. 

\section{Data informativity for robust stabilization}
\label{sec:data-informativity}
We are now ready to introduce the fundamental characterization  of the stabilizability condition for the systems consistent with the data according to Definition~\ref{defn:consistency}. 
We begin by focusing on the stabilization of a single class $\Sigma^v$ by means of state feedback~\cite{van2020data}.
\begin{definition}
    The dataset $\mathcal{D}^v$ is called informative for stabilization by state feedback if there exists $K \in \mathbb{R}^{m \times n}$ such that the closed-loop system 
        $(A + BK)$
    is Schur stable for any $(A,B)$ in $\Sigma^v$.
\end{definition}

The condition expressed by the above definition is equivalent to the existence of a quadratic Lyapunov function $V(x) = x^T P x$ with $P \in \mathbb{R}^{n \times n}$, $P = P^T \succ 0$, and a gain $K$ for which
\begin{equation}\label{eq:Lyap_stabiliz}
P - (A + BK)^T P (A + BK) \succ 0
\end{equation}
holds for all $(A,B)\in\Sigma^v$.
This can be posed as an LMI through the change of variables $P = \gamma H^{-1}$ and $K = LH^{-1}$, with $L \in \mathbb{R}^{n \times m}$ \cite{boyd1994linear}. 
Thus,
\eqref{eq:Lyap_stabiliz} is equivalent to 
\begin{equation}\label{eq:LMI-feasibility-single}
\begin{bmatrix}
    I \\ A^T \\ B^T
\end{bmatrix}^T
\underbrace{
\begin{bmatrix}
        H & 0 & 0 \\
        0 & -H &-L^T  \\
        0 & -L &- L H^{-1} L 
    \end{bmatrix}}_{M}
    \begin{bmatrix}
    I \\  A^T \\ B^T
\end{bmatrix} \succ 0.
\end{equation}

By \eqref{eq:consistency-characterization}, for all $(A,B) \in \Sigma^v$ we have
\begin{equation}\label{eq:data-consistent-systems}
\begin{bmatrix}
    I \\ A^T \\ B^T
\end{bmatrix}^T
\underbrace
{\left(\begin{bmatrix}
    {X}^{v}_+ \\-{X}^{v}_-\\-{U}^{v}_-
\end{bmatrix}
\begin{bmatrix}
    {X}^{v}_+ \\-{X}^{v}_-\\-{U}^{v}_-
\end{bmatrix}^T
\right)}_{-N^{v}}
\begin{bmatrix}
    I \\ A^T \\ B^T
\end{bmatrix} = 0.
\end{equation}
Finally, \eqref{eq:LMI-feasibility-single} can be associated with the class $\Sigma^v$ through Finsler's lemma \cite{van2021matrix}.
Thus, a stabilizing data-driven controller can be characterized by finding $H \succ 0, L,$ and $\epsilon^v >0$ such that
\begin{equation}\label{eq:LMI-finsler}
    M - \epsilon^v N^{v} \succ 0,
\end{equation}
where $M$ and $N^v$ are defined in \eqref{eq:LMI-feasibility-single} and \eqref{eq:data-consistent-systems}, respectively. 
In fact, for any pair $(H,L)$ fulfilling~\eqref{eq:LMI-finsler}, one can retrieve a feedback $u_k = K x_k$ which stabilizes all linear systems compatible with the data $\setD^v$, i.e., all systems defined by $(A,B)\in\Sigma^v$, and a common Lyapunov function defined by the matrix $P$ \cite{van2021matrix}. 

This result readily extends to LDI systems as follows.
\begin{lemma}\label{lem:robustLDI}
Let the matrix inequalities
\begin{multline}\label{eq:Finsler_vertices}
\underbrace{
\begin{bmatrix}
        H & 0 & 0 &0 \\
        0 & -H &-L^T & 0  \\
        0 & -L &0 & L  \\
        0 & 0 &L^T & H \\
    \end{bmatrix}}_{\widetilde{M}}
    - \epsilon^v
\underbrace{\left(-
\begin{bmatrix}
    {X}^{v}_+ \\-{X}^{v}_-\\-{U}^{v}_- \\ 0
\end{bmatrix}
\begin{bmatrix}
    {X}^{v}_+ \\-{X}^{v}_-\\-{U}^{v}_- \\ 0
\end{bmatrix}^T\right)}_{\widetilde{N}^v}
\succ 0,\\
v = 1,\ldots,{n_v},
\end{multline}
be satisfied for some $H\in\mathbb{R}^{n\times n}$, $H = H^T \succ 0$, $L\in\mathbb{R}^{m\times n}$, and scalars $\epsilon^1,\epsilon^2,\ldots,\epsilon^{n_v} > 0$, for the data $({X}^{v}_{+},{X}^{v}_{-},{U}^{v}_{-})$ in the collection of vertex datasets $\mathcal{D} = \{\mathcal{D}^v\}_{v=1}^{n_v}$. Then, the LDI \eqref{eq:sys_lpv} is robustly stabilized by the linear feedback $K=LH^{-1}$, and the matrix $P=H^{-1}$ defines a Lyapunov function for the associated closed-loop LDI dynamics.
\end{lemma}
\begin{proof}
    First, note that \eqref{eq:LMI-finsler} can be cast as \eqref{eq:Finsler_vertices} by application of the Schur complement. It can be verified that $\widetilde{M}$ and $\widetilde{N}^v$, for $v=1,\ldots,{n_v}$, satisfy Finsler's lemma's assumptions by construction~\cite{van2021matrix}. Then, satisfaction of \eqref{eq:Finsler_vertices} implies that any linear system described by $(A,B)\in\SigmaV$ is stabilized by $K$. Since additional data remove degrees of freedom, $\SigmaV\supseteq\mathcal{V}$ for any nonempty set of data. Then, \eqref{eq:Finsler_vertices} is equivalent to $P - (A + BK)^T P (A + BK) \succ 0$ for all $(A,B)\in\textit{Co}(\mathcal{V})$ (see, e.g.,~\cite[\S5.1]{boyd1994linear}).
\end{proof}

Lemma~\ref{lem:robustLDI} yields the following definition of data informativity for 
robust state feedback.
\begin{definition}
\label{defn:informativity_robust}
    A dataset $\{D^v\}_{v=1}^{n_v}$ is \emph{informative for LDI} stabilization if there exists $K \in \mathbb{R}^{m \times n}$ and $P\in\mathbb{R}^{n\times n}$, $P = P^T \succ 0$, such that $P-(A+BK)^T P(A+BK) \succ 0$ for all $(A,B)\in\SigmaV$.
\end{definition}
In the following,
we designate by $\setKD$ the set of all pairs $\KP$ which solve \eqref{eq:Finsler_vertices}.

\section{Data-consistent Optimal Control}\label{sec:data-informativity-lqr}
Considering a given dataset $\mathcal{D}^o$ which determines the class $\Sigma^o$, the following optimal control problem can be posed:
\begin{equation}\label{eq:optimal_bound}
    \min_{K}   J_{cl}(x_k, K, \Sigma^o),
\end{equation}
where
\begin{equation}\label{eq:online_cost-to-go}
    \begin{split}
     J_{cl}(x_k, K, \Sigma^o) \coloneqq{}& \max_{A^o_j,B^o_j} \, \sum_{j=0}^\infty \bar{x}^T_{j} (Q + K^T R K)\bar{x}_j,\\ 
    \mathrm{s.t.} \quad \bar{x}_{j+1} ={}& (A^o_j + B^o_j K)\bar{x}_j,\; \bar{x}_0 = x_k,\\
    (A^o_j,B^o_j)&\in\Sigma^o,\; j = 0,1,\ldots
\end{split} 
\end{equation}

First, note that feasibility of \eqref{eq:optimal_bound} implies the existence of a finite $J_{cl}(x_k, K, \Sigma^o)$
for some stabilizing $K$.
By virtue of \cite{kothare1996robust}, we can find a candidate Lyapunov function $V(x_k) \coloneqq x_k^T P x_k $,  with $P = P^T \succ 0$, such that
$V(x_k) \geq J_{cl}(x_k, K, \Sigma^o)$. 
For this, we can state the stabilizing condition $V(x_k)-V(x_{k+1}) \geq x_k^T(Q+K^TRK)x_k$, which leads to the following inequality:
\begin{equation}\label{eq:perf_lower_bound1}
    P - (A+BK)^T P (A+BK) \succeq Q + K^T R K,
\end{equation} 
for all ${(A,B)\in \Sigma^o}$.
Through the change of variables $P = \gamma H^{-1}$ and $K = LH^{-1}$, with $L \in \mathbb{R}^{n \times m}$, pre- and post-multiplying \eqref{eq:perf_lower_bound1} by $H$, we get
\begin{equation}\label{eq:perf_lower_bound2}
    H -(AH+BL)^T H^{-1}(AH+BL)-\gamma^{-1} \Phi^T \Phi \succeq 0,
\end{equation}
where $\Phi := \begin{bmatrix}(\hat{Q}H)^T & (\hat{R}L)^T\end{bmatrix}^T$, with $\hat{R}^T \hat{R} = R$, $\hat{Q}^T \hat{Q} = Q$.
Following~\cite{nguyen2023lmi}, we obtain the equivalent conditions
\begin{equation}\label{eq:M_Phi}
    \begin{bmatrix}
    I \\ A^T \\ B^T
\end{bmatrix} ^T 
    \underbrace{
    \begin{bmatrix}
        H & 0 \\
        0 & \begin{bmatrix}
            H \\ L
        \end{bmatrix} 
         (H - \gamma \Phi^T \Phi) ^{-1}
         \begin{bmatrix}
            H \\ L
        \end{bmatrix}^T
    \end{bmatrix}}_{M_\Phi}
    \begin{bmatrix}
    I \\ A^T \\ B^T
\end{bmatrix} \succeq 0.
\end{equation}
\noindent The data-consistency condition (cf.~\eqref{eq:data-consistent-systems}) with respect to the online data is given by
\begin{equation}
    \begin{bmatrix}
    I \\ A^T \\ B^T
\end{bmatrix} ^T \underbrace{\left(
 \begin{bmatrix}
    {X}^{o}_+ \\-{X}^{o}_-\\-{U}^{o}_- 
\end{bmatrix}
\begin{bmatrix}
    {X}^{o}_+ \\-{X}^{o}_-\\-{U}^{o}_-
\end{bmatrix}^T\right)}_{-{N}^o_\Phi}
    \begin{bmatrix}
    I \\ A^T \\ B^T
\end{bmatrix} = 0.
\label{eq:N_Phi}
    \end{equation}

Now, by Finsler's lemma, equation~\eqref{eq:M_Phi} is fulfilled for all $(A,B)$ such that~\eqref{eq:N_Phi} holds if and only if there exists $\epsilon^o>0$ such that
\begin{equation*}
    M_\Phi - \epsilon^o {N}^o_\Phi \succeq 0.
\end{equation*}
This condition can be equivalently stated as an LMI through the application of Schur complement:
\begin{subequations}
\label{eq:LMI-informativity-LQR}
\begin{equation}
\label{eq:M_Phi_positive}
    \begin{bmatrix}
        H & 0 & 0 &0 &0\\
        0 & 0 &0 & H &0 \\
        0 & 0 &0 & L &0 \\
        0 & H &L^T & H & \Phi^T \\
        0 & 0 &0 & \Phi &\gamma I \\
    \end{bmatrix} 
    - \epsilon^o 
 \left(-\begin{bmatrix}
    {X}^{o}_+ \\-{X}^{o}_-\\-{U}^{o}_- \\ 0 \\ 0
\end{bmatrix}
\begin{bmatrix}
    {X}^{o}_+ \\-{X}^{o}_-\\-{U}^{o}_- \\ 0 \\ 0
\end{bmatrix}^T\right)\succeq 0.
\end{equation}
    \begin{equation}\label{eq:schur-b-lqr}
    \begin{bmatrix}
        H & \Phi \\ \Phi & \gamma I
    \end{bmatrix} \succeq 0.
\end{equation}
\end{subequations}

\begin{lemma}\label{lem:optcontrol}
    Let the dataset $\mathcal{D}^o$ be such that~\eqref{eq:LMI-informativity-LQR} holds for some $H, L, \epsilon^o$ and $\gamma$. Then for all possible sequences $\{(A^o_j,B^o_j) \in \Sigma^o\}_{j=0}^\infty $ the feedback $K = LH^{-1}$ is stabilizing and $V(x) = x^T P x$ with $P = \gamma H^{-1}$ is a common Lyapunov function for all the data-consistent closed-loop dynamics.
    Also, $V(x_k)$ is an upper bound for $J_{cl}(x_k,K,\Sigma^o)$.
\end{lemma}
\begin{proof}
    Note that the matrices in~\eqref{eq:LMI-informativity-LQR} fulfill the conditions in Finsler's lemma by construction~\cite{van2021matrix}.
    This implies that~\eqref{eq:M_Phi} holds for all $(A^o,B^o) \in \Sigma^o$ and, by following the reasoning~\eqref{eq:perf_lower_bound1} to~\eqref{eq:LMI-informativity-LQR} in reverse order, we have that~\eqref{eq:perf_lower_bound1} holds for the system $(A_k,B_k) \in \Sigma^o$ that maximizes $V(x_{k+1})$ and, therefore, the inequality holds for all $(A^o,B^o) \in \Sigma^o$.
    This proves that $V(x_k)$ is a Lyapunov function and the asymptotic stability is explained for any realization of $\{(A^o_k,B^o_k)\}_{k=0}^\infty$, $(A^o_k,B^o_k) \in \Sigma^o$. 
    Moreover, evaluating
    \begin{displaymath}
     \sum_{k=0}^\infty V(x_k)-V(x_{k+1}) 
     \geq \max_{(A_k,B_k)\in\Sigma^o}\sum_{k=0}^\infty x_k^T (Q + K^T R K)x_k
    \end{displaymath}
    subject to ${x}_{k+1} = (A_k + B_k K){x}_k$, all inner terms cancel out in the left hand-side and $V(x_\infty) = 0$ since the closed-loop system is stable.
    It follows that $V(x_k) \geq J_{cl}(x_k,K,\Sigma^o)$.
\end{proof}
\begin{remark}
The solution to the minimization problem $\min_{H,L,\gamma>0} \gamma$ subject to \eqref{eq:LMI-informativity-LQR} determines the minimum upper bound on the cost functional for the optimal feedback gain $K=LH^{-1}$.
    In the case $\Sigma^o = \{(A^o,B^o)\}$, this minimization results in a data-driven LQR problem. 
\end{remark}

\section{Adaptive Data-Driven Control for Systems under Polytopic Parametric Uncertainty}
\label{sec:robust}
In this section we introduce the main contribution of this work. 
As introduced earlier in the text, we seek to optimize the control performance consistently with the online collected data, without compromising stability for the LDI dynamics \eqref{eq:sys_lpv}.
We consider the following blanket assumption.
\begin{assumption}\label{assum:feas}
The dataset $\setD = \{\setD^v\}_{v=1}^{n_v}$ is informative for LDI stabilization by state feedback as per Definition~\ref{defn:informativity_robust}.
\end{assumption}
We emphasize that any vertex dataset $\setD^v$ can comprise triplets extracted from separate experiments. This implies that, in practice, a dataset $\setD$ fulfilling Assumption~\ref{assum:feas} can be obtained by gathering a sufficient number of independent triplets corresponding to the system operating at some vertex.  The satisfaction of Assumption~\ref{assum:feas} can be verified offline. 

From now on, we use the subscript $k$ to distinguish the online dataset updated at time $k$, i.e., $\mathcal{D}^o_k$. Similarly, we refer to the class of systems consistent with it as $\Sigma^o_k$.
Then, the logic behind our approach can be summarised through the following optimization problem:\footnote{In the remainder, we 
parametrize the Lyapunov function $V$ by the pair $(K,P)$ to avoid ambiguities.}
\begin{subequations}\label{eq:adaptive_orig}
\begin{align}
  &(K_k,P_k) \in \argmin_{K,P}  V(x_k, \KP)\label{eq:adaptive_cost_orig}\\
    \mathrm{s.t.} \quad
    & J_{cl}(x_k, K, \Sigma^o_k) \leq V(x_k, \KP), \label{eq:cotaV}\\
    & V(\mathbb{S}(x_k,K x_k,\SigmaV),(K,P)) < V( x_k,\KP),\label{eq:decreas_Lyap_LDI}\\
    &{\varXi} \subseteq \mathcal{X},\; K {\varXi} \subseteq \mathcal{U}. \label{eq:constr_orig}
\end{align}
\end{subequations}
where ${\varXi} \coloneq \{x\colon V(x,\KP) \leq \gamma \}$, $\gamma = V(x_k,\KP)$  (i.e., the sublevel sets of $V(x_k,\KP)$); in~\eqref{eq:decreas_Lyap_LDI} we overload the notation with set-valued arguments for brevity and, in agreement with our data-driven approach, we express the vertex systems $\mathcal{V}$ through the class of systems consistent with the offline data $\SigmaV$.

The objective of \eqref{eq:adaptive_orig} is to find a feedback, parametrized by $(K_k,P_k)$, that minimizes the Lyapunov function $V(x_k,(K_k,P_k))$---which by~\eqref{eq:cotaV}  stands as an upper bound of $J_{cl}(x_k, K_k, \Sigma^o_k) $ (cf.~Section \ref{sec:data-informativity-lqr}). 
The same feedback must guarantee contractivity of $V$ for any possible realization of the closed-loop LDI dynamics~\eqref{eq:decreas_Lyap_LDI} (cf.~Section \ref{sec:data-informativity}).
Finally,~\eqref{eq:decreas_Lyap_LDI} together with \eqref{eq:constr_orig} ensure invariance of~$\varXi$ under admissible inputs.
We translate~\eqref{eq:constr_orig} into a tractable form as follows.
From \cite{boyd1994linear}, it is known that
an ellipsoid $\mathcal{E}(\gamma) = \{x \in \mathbb{R}^n : x^TPx \leq \gamma\}$ is contained in the set $\mathcal{W} = \{x\in \mathbb{R}^n: w_i x \leq 1, i = 1, \dots, r\}$ if and only if $w_i (\gamma P^{-1})w_i^T \leq 1, i = 1,\dots,r)$. 
 Expressing~\eqref{eq:constraints} by rows we have $w^i_x x \leq 1$ and $w^j_u u\leq 1$ (i.e., $w^j_u Kx\leq 1$), which lead to $w^i_x (\gamma P^{-1})(w^i_x)^T \leq 1$ and $(w^j_u K)(\gamma P^{-1})(w^j_u K)^T \leq 1,$ for $i = 1,\dots,r$ and $ j = 1,\dots,\ell$.
 Recalling the change of variables $P = \gamma H^{-1}$, $K = LH^{-1}$ 
 the constraints are arranged in LMI form as follows%
\begin{subequations}\label{eq:LMI-x-u-constraints}
\begin{align}
    &\begin{bmatrix}
        1 & x_k^T \\ x_k & H
    \end{bmatrix} \succ 0,\label{eq:invX} \\
&\begin{bmatrix}
        1 & w^i_x H \\ (w^i_x H)^T & H
    \end{bmatrix} \succ 0, \quad i = 1,\dots,r,\label{eq:constrXlmi}\\
    &\begin{bmatrix}
        1 & w^j_u L \\ ( w^j_u L)^T & H
    \end{bmatrix} \succ 0,  
     \quad j = 1,\dots, \ell.\label{eq:constrUlmi}
\end{align}
\end{subequations}

Then, by recasting \eqref{eq:adaptive_cost_orig} into an epigraphic form, and using Lemmas~\ref{lem:robustLDI} and~\ref{lem:optcontrol}, the adaptive control law is obtained as the solution of
\begin{equation}\label{eq:LMI_DATA}
\begin{split}
  \probLMI{\SigmaV}{\Sigma^o_k}: \quad &   \min_{H,L, \gamma, {\epsilon}^v, \epsilon^o} 
    \gamma \\
    \mathrm{s.t.} \quad &
\eqref{eq:Finsler_vertices},\;\eqref{eq:LMI-informativity-LQR},\;\eqref{eq:LMI-x-u-constraints},\\
&\gamma, \epsilon^o,{\epsilon}^v > 0,\\
&v =1,\ldots, n_v.
\end{split}
\end{equation}

The main properties of the proposed controller are presented below. We rely on the following assumption, discussing its relaxation in the last part of this section.

\begin{assumption}\label{assum:onlinedata}
    At each time $k$, the online data $\Sigma^o_k$ is such that a single system is consistent with it, i.e., $\Sigma^o_k = \{(A^o_k,B^o_k)\}$. Moreover,  $(A^o_k,B^o_k)\in\mathit{Co}(\SigmaV)$.
\end{assumption}
The first part of Assumption~\ref{assum:onlinedata} is equivalent to having
    $\mathit{rank}\begin{bmatrix}
        X^o_-\\
        U^o_-
    \end{bmatrix} = n+m$,
which is known as \emph{persistent excitation} or \emph{informativity for system identification}; see \cite{van2020data,de2019formulas}. 
\begin{remark}

Assumption~\ref{assum:onlinedata} is very restrictive in practice, as it hardly fits the considered time-varying LDI dynamics. Its purpose is merely that of simplifying the technical discussion in the sequel. We analyse its relaxation---mainly aimed at piece-wise constant parametric variations---in Section~\ref{sec:assum2}.
\end{remark}
\begin{thm}\label{thm:main_result}
     Let Assumptions~\ref{assum:feas} and~\ref{assum:onlinedata} hold. Take  $x_k$ and $\Sigma^o_k$ such that~$\probLMI{\SigmaV}{\Sigma^o_k}$ is feasible. 
     Then, the receding-horizon feedback law $u_k = K_k x_k$ with $K_k = L H^{-1}$, where $H,L$ are obtained as the solution of~\eqref{eq:LMI_DATA} is such that
         \textit{i)} \eqref{eq:LMI_DATA} is recursively feasible, 
         \textit{ii)} the closed-loop system is robustly stable, and
         \textit{iii)} if there exists  $\bar{k}$ such that $\Sigma^o_{k+1} = \Sigma^o_{k}$ for all $k \geq \bar{k}$ then the closed loop achieves the minimal infinite-horizon cost \eqref{eq:online_cost-to-go} w.r.t.~the 
         feasible stabilizing feedbacks in~$\setKD$.
\end{thm}
\begin{proof}
\textit{i)} To prove recursive feasibility, it suffices to show that the availability of a solution of $\probLMI{\SigmaV}{\Sigma^o_k}$ guarantees the satisfaction of \eqref{eq:cotaV},  \eqref{eq:decreas_Lyap_LDI} and \eqref{eq:constr_orig} for any $(x_{k+1}, \Sigma^o_{k+1})$. 
Let $\gamma_k, P_k = \gamma_k H_k^{-1}, K_k = L_k H_k^{-1}$ be obtained from~\eqref{eq:LMI_DATA} at time $k$. 
By \eqref{eq:invX} and \eqref{eq:constrXlmi}, $x_k\in\varXi_k = \{x\colon x^T P_k x\leq \gamma_k\} = \{x\colon V(x,\KPk) \leq V(x_k,\KPk)\}\subseteq \mathcal{X}$; \eqref{eq:constrUlmi} guarantees that $u=K_k x\in\mathcal{U}$ for all $x\in\varXi_k$. Finally, \eqref{eq:Finsler_vertices} implies that $x_{k+1}$ must belong to the interior of $\varXi_k$ (cf.~Lemma~\ref{lem:robustLDI}). Hence, \eqref{eq:constr_orig} is fulfilled at $k+1$.
From the same Lemma~\ref{lem:robustLDI}, we have that \eqref{eq:decreas_Lyap_LDI} is also verified for any $x_{k+1}\in\mathbb{S}(x_k,K x_k,\SigmaV)$.

It remains to prove feasibility of \eqref{eq:cotaV}. 
Recall that, by Assumption~\ref{assum:onlinedata}, $\Sigma^o_{k+1} \subseteq \mathit{Co}(\SigmaV)$: hence, it follows from \eqref{eq:online_cost-to-go} that $J_{cl}(x_{k+1}, K_k, \Sigma^o_{k+1})\leq J_{cl}(x_{k+1}, K_k, \mathit{Co}(\SigmaV))\leq \alpha  V(x_{k+1},(K_k,P_k))$ for some $\alpha > 0$.
Then, we can define $\hat{K}=K_k$, $\hat{P} = \alpha P_k$ and $\hat{\gamma} = \alpha\gamma_k$, noticing that by construction \eqref{eq:decreas_Lyap_LDI} and \eqref{eq:constr_orig} still hold for $K = \hat{K}$, $P = \hat{P}$ and $\gamma = \hat{\gamma}$.
Therefore, $J_{cl}(x_{k+1},K_k,\Sigma^o_{k+1}) \leq  V(x_{k+1},(\hat{K},\hat{P}))$, and \eqref{eq:adaptive_orig} is recursively feasible with $K_{k+1} = \hat{K}$, $P_{k+1} = \hat{P}$, $\gamma_{k+1} = \hat{\gamma}$; for the equivalent formulation \eqref{eq:LMI_DATA}, note that the satisfaction of the above conditions also implies the existence of feasible $\epsilon^o_{k+1},{\epsilon}^v_{k+1} > 0$.

\textit{ii)} Once recursive satisfaction of \eqref{eq:Finsler_vertices} in \eqref{eq:LMI_DATA} is established, robust stability follows directly from Lemma~\ref{lem:robustLDI}. 
\textit{iii)} We are left to discuss the online performance improvement. Assumption~\ref{assum:onlinedata} 
together with the stated hypothesis imply that
\eqref{eq:sys_lpv} operates about a region described by some linear model $(A^o,B^o)\in\mathit{Co}(\SigmaV)$, for all $k\geq \bar{k}$. 
Recall that \eqref{eq:LMI-informativity-LQR} expresses \eqref{eq:cotaV} by virtue of Lemma~\ref{lem:optcontrol}.
Then, for all $k\geq \bar{k}$, $V(x_k, \KPk)$ is an upper bound of the cost-to-go of \eqref{eq:sys_lpv} in closed-loop, i.e., 
\begin{equation}\label{eq:cotaV_proof}
 \sum_{j=0}^\infty \bar{x}^T_{j} (Q + K^T R K)\bar{x}_{j} \leq x_k^T P x_k,   
\end{equation}
where $\bar{x}_{j+1}\in\mathbb{S}(\bar{x}_j,K \bar{x}_j,\{(A^o, B^o)\}) = (A^o+B^o K)\bar{x}_j$, for all $j\geq 0$, and $\bar{x}_0 = x_k$. Since $K,P$ are decision variables (bound to $\setKD$ to satisfy \eqref{eq:decreas_Lyap_LDI}, and restricted by \eqref{eq:constr_orig}), with $P$ arbitrarily scalable due to the reasons discussed in part \textit{i)} of this proof, the inequality in \eqref{eq:cotaV_proof} always holds with equality at solution, and its LHS is minimized accordingly through \eqref{eq:adaptive_cost_orig}. This proves the final claim and concludes the proof.
\end{proof}

\subsection{Relaxing Assumption~\ref{assum:onlinedata}}
\label{sec:assum2}
It is possible to relax Assumption~\ref{assum:onlinedata}.
In such a case, a solution of $\probLMI{\SigmaV}{\Sigma^o_k}$ (cf.~\eqref{eq:LMI_DATA}) might not exist.  Nonetheless, stability can still be guaranteed as follows (as also illustrated in the numerical example of Section~\ref{sec:numerical}). We first analyse the case where \eqref{eq:LMI_DATA} is feasible. If $\Sigma^o_k \subseteq \SigmaV$, 
then the properties of the controller correspond to those stated in Theorem~\ref{thm:main_result}.
Otherwise, $\Sigma^o_k \not\subseteq \SigmaV$: in this case, $\Sigma^o_k$ may or may not be a singleton. If it is, then 
the pair $\KPk$ obtained as solution of \eqref{eq:LMI_DATA} defines a stabilizing feedback law for a new LDI $\mathbb{S}(x,u,\mathcal{V}')$, where $\mathcal{V}' = \SigmaV\cup \{(A^o,B^o)\}$. Since $\mathit{Co}(\mathcal{V}')\supset\SigmaV$, the closed loop will be stable for the original system \eqref{eq:sys_lpv}. A similar analysis applies for the case where $\Sigma^o_k$ is multi-valued. In such a case, it is worth pointing out that $\mathit{Co}(\mathcal{V}')$, where $\mathcal{V}' = \SigmaV\cup\Sigma^o_k$, might even be unbounded in the space of system matrices $\mathbb{R}^{n\times n} \times \mathbb{R}^{n\times m}$, possibly expressing degenerate system realizations; we refer the reader to \cite[\S IV]{van2020data} for a detailed discussion on this aspect.

It remains to discuss the case where \eqref{eq:LMI_DATA} is infeasible at some time $k'>k$.
When $\Sigma^o_{k'} \neq \emptyset$, this is due to the impossibility of finding a pair $\KP\in\mathcal{K}_{\mathcal{D}}\cap\mathcal{K}_{\mathcal{D}^o}$ which satisfies the problem's constraints. If instead $\Sigma^o_k = \emptyset$, constraint~\eqref{eq:LMI-informativity-LQR} becomes infeasible due to loss of rank of $\tilde{N}^o_\Phi$. Under either of these circumstances, the closed loop can be maintained by using the previously obtained feedback $K_{k-1}$: robust stability is still guaranteed by the invariant defined by $\Gamma_{k-1}$. Alternatively, \eqref{eq:LMI_DATA} can be solved by using the previous dataset $\mathcal{D}^o_{k-1}$, i.e., by considering the problem $\probLMI{\SigmaV}{\Sigma^o_{k-1}}$. In this case, a solution is guaranteed to exist owing to the recursive feasibility of \eqref{eq:LMI_DATA}; moreover, by contractivity of the LDI trajectory (which follows from \eqref{eq:LMI-finsler}), the newly found $(K_k,P_k)$ results in a feedback action no more conservative than $(K_{k-1},P_{k-1})$ due to~\eqref{eq:LMI-x-u-constraints}.

\section{Numerical evaluation}\label{sec:numerical}
In this section, we illustrate the effectiveness of the proposed controller through numerical simulations. 
As benchmark, we recall the numerical example from~\cite{kothare1996robust}, which models an angular positioning system,
\begin{equation*}
    x_{k+1} = \begin{bmatrix}
        1 & 0.1 \\ 0 & 1- 0.1 \delta_k
    \end{bmatrix} x_{k} + \begin{bmatrix}
        0 \\ 0.1 \rho
    \end{bmatrix}
    u_k,
\end{equation*}
where $\rho = 7.87$ and $0.1 \leq \delta_k \leq 10$, and the input is constrained as $-1 \leq u_k \leq 1$, $k\geq 0$. In this case, the LDI is configured by the vertex systems corresponding to $\delta_k = 0.1$ and $\delta_k = 10$, respectively.
The vertex datasets are obtained by applying random input sequences of length $T^{1} = T^2 = 3$ (which satisfies Assumption~\ref{assum:feas}) to each vertex system from the initial state $x_0 = [0.95,0]^T$.
The online data window size is chosen as $T^o = 3$. 

We define the performance metric $J \coloneqq \sum_{k=T^o}^{T_\mathrm{sim}} x_k^T Q x_k + u_k^T R u_k$, where $Q = \mathit{diag}([1,1])$, $R = 0.01$ and $T_\mathrm{sim} = 50$ steps. This is used to compare the performance of the proposed adaptive controller (designated by $J_A$) with that of its robust counterpart ($J_R$); the latter is obtained by enforcing~\eqref{eq:cotaV} over $\SigmaV$ in the spirit of \cite{kothare1996robust} (see \cite{nguyen2023lmi} for a related data-driven reformulation). 
All simulations are implemented in MATLAB through YALMIP~\cite{lofberg2004yalmip}, using the MOSEK solver~\cite{mosek}. Note that, as indicated in Section~\ref{sec:assum2}, the previously computed feedback is maintained whenever the online data is not sufficiently informative.
 
Figure~\ref{fig:simulation} illustrates the results obtained for the case $\delta_k = 2$, $k=0,\ldots,14$, and $\delta_k = 4$, $k\geq 15$.
Note that the online data sequence for the adaptive case is initialized by controlling the system by the robust-only formulation during the first $T^o$ time steps (after which the adaptive controller is engaged).
The adaptive and robust performance metrics are $J_A = 8.651$ and $J_R = 8.8794$, respectively, with the adaptive cost being $2.6$\% lower.

\begin{figure}[t]
    \centering    
    \includestandalone{figurita}
    \vspace{-2em}
    \caption{States and input trajectories for the adaptive (orange) and robust (blue) controllers (the robust controller is used for both cases during the first $T^o = 3$ time steps). The value of $\delta_k$ is set to $2$ for $k=0,\ldots,14$, then $\delta_k = 4$ for $k\geq 15$. The resulting performances are $J_A = 8.651$ and $J_R = 8.8794$, with the adaptive controller cost being $2.6$\% lower.}
    \label{fig:simulation}
\end{figure}

We performed a separate series of simulations where $\delta_k = 2$ for $k = 0,\ldots,14$, and $\delta_k =\bar{\delta}$ for $k\geq 15$; the value $\bar{\delta}$ is sampled from the uniform distribution over $[0.1, 10]$, for a total of 150 experiments. In all cases, the same initial state $x_0 = [0.95,0]^T$ is used.
Figure~\ref{fig:performance} shows the performance gain $(J_A - J_R)/J_R$ obtained by using the proposed adaptive formulation \eqref{eq:LMI_DATA} over its robust counterpart: it can be observed that the closed-loop cost is reduced by $2.5$--$5\%$ in all instances.

\begin{figure}[t]
    \input{boxplot_minus.tikz}
    \vspace{-2em}
    \caption{Cost reduction ${(J_A - J_R)}/{J_R}$ for 150 experiments, where $\delta_k = 2$, $k=0,\ldots,14$ and $\delta_k = \bar{\delta}, k\geq 15$; the box plots summarize the results for 15 different values for $\bar{\delta}$ sampled uniformly from each of the subintervals of $[0.1, 10]$ indicated on the horizontal axis.}
    \label{fig:performance}
\end{figure}

\section{Conclusions}\label{sec:discussion}
We presented an adaptive data-driven predictive controller for LDI dynamics. 
The design relies on offline data collected from a finite number of experiments, together with updated state and input trajectories collected online in the form of a fixed length sliding window.
The controller formulation relies on the data informativity notion. When the data is sufficiently informative for identification, we show how the control law can adapt to the online operating condition of the system.
The properties of the proposed controller with respect to non-adaptive formulations are illustrated in simulation. 
Future work includes extending the adaptive approach to allow for 
perturbations in the data---e.g., additive measurement noise or small  parametric variations---thereby widening the context of application, currently limited by Assumption~\ref{assum:onlinedata}.

\vspace*{-2pt}
\section*{Acknowledgments}
\vspace*{-1pt}
\noindent The authors would like to thank Dr.~Gerson Portilla and Prof.~Teodoro Alamo for the enlightening discussions. We are also grateful to the anonymous reviewers for their valuable contribution to the quality of the manuscript.

\bibliography{data-driven-ldi}

\end{document}